\documentclass[11pt,a4]{article}

\usepackage{setspace}
\usepackage{geometry}
\usepackage{rotating, graphicx}
\usepackage{amsthm,amsmath,latexsym,amssymb,amsfonts,amscd}
\usepackage{graphics,lscape,fancyhdr,array,stmaryrd,euscript}
\usepackage{empheq}
\usepackage{dsfont}
\usepackage{verbatim}
\usepackage{color,tikz,tikz-cd}
\usetikzlibrary[snakes]
\usepackage{relsize,slashed}
\numberwithin{equation}{section}
\usepackage{hyperref}
\usepackage{setspace}

 \usepackage[numbers,sort&compress]{natbib}
 
 \usepackage[nottoc]{tocbibind}
\date{}

\begin{document}
\hfill
\vspace{-1.5cm}
\vskip 0.05\textheight
\begin{center}
{\Large\bfseries 
Holography of pp-waves in conformal gravity}

\vspace{0.4cm}

\vskip 0.03\textheight
Ansh Bhatnagar${}^{a}$,
Iva Lovrekovic${}^b$

\vskip 0.03\textheight

\vspace{5pt}
{\em
$^a$ Theoretical physics group, Blackett Laboratory,\\ Imperial College London, SW7 2AZ, U.K.\\
\vspace{5pt}

$^b$
Technische Universit\"at Wien,\\
Wiedner Haupt. 8-10, 1040, Wien, Austria \\
}

\abstract{We consider holography of two pp-wave metrics in conformal gravity, their one point functions, and asymptotic symmetries. One of the metrics is a generalization of the standard pp-waves in Einstein gravity to conformal gravity. The holography of this metric shows that within conformal gravity one can have realised solution which has non-vanishing partially massless response (PMR) tensor even for vanishing subleading term in the Fefferman-Graham expansion (i.e. Neumann boundary conditions), and vice-versa.  } 
\end{center}



\section{Introduction}
Conformal gravity is a higher derivative theory of gravity which has its recurrent appearance in the literature. It is power-counting renormalizable and highly symmetric which makes it interesting for studying \cite{Mannheim:2011ds,Hooft:2010ac}. Main argument against the theory is its non-unitarity, which manifests for example in two-point correlation functions \cite{Ghodsi:2014hua}. That issue is addressed via known methods \cite{Mannheim:2011ds,Bender:2007wu}, or the theory is considered as a toy model for its symmetry properties. From phenomenological aspects CG explains galactic rotation curves without addition of dark matteer \cite{Mannheim:1988dj}, and it was also stated to be an exact solution to perturbative cosmology in recombination era \cite{Mannheim:2020lfz}.  
The analysis of the asymptotic symmetries of CG in 3+1 dimensions allows for classification of the asymptotic solutions \cite{Irakleidou:2016xot}. There is no classification of the global cosmological solutions in CG, however number of Einstein gravity (EG) solutions have been generalized to CG \cite{Liu:2012xn,Starobinsky:1982mr}.  Four dimensional cosmological solutions of Einstein gravity (EG), have of course been most studied \cite{Stephani:2003tm}, and most classified. Most popular classifications to date are Bianchi classification, and Petrov \cite{Petrov:2000bs} classification where the latter often uses Newman-Penrose formalism. 
Here, we calculate two general solutions of the pp-wave metric with and without a cylinder in CG and analyse their asymptotics. 

CG holography has in the earlier studies showed that in the framework of AdS/CFT, there are two holographic stress energy tensors at the boundary. One of them analogous to Brown-York (BY) stress energy tensor and another called partially massless response (PMR) which does not have an analog in EG. Holographic analyses of Schwarzschild solution in EG, Manheimm-Kazanas-Riegert solution in CG \cite{Mannheim:1988dj}, and rotating black hole solution in AdS with Rindler hair \cite{Liu:2012xn}, showed that their PMR vanishes when generalized Fefferman-Graham boundary conditions reduce to standard Fefferman-Graham boundary conditions used in EG. \footnote{Generalized Fefferman-Graham boundary conditions allow for the subleading term in the expanision in hologrpahic coordinate, around the boundary of the manifold. In standard Fefferman-Graham expansion this term is set to zero \cite{Starobinsky:1982mr}.}
The pp-wave solutions which we analyse here, show that it is possible to have vanishing PMR for the generalized Fefferman-Graham (FG) boundary conditions and that it is possible to have non-vanishing PMR for standard FG boundary conditions.
Vanishing of the PMR, also implies vanishing of the corresponding two point correlation functions.
Beside the holography of the solution, we consider 
its Killing vectors, charges, and asymmptotic symmetry algebra (ASA) 
as well as speculate possibility of using the  metric as a cosmological background for the string quantization.

\section{Conformal gravity}
Given a manifold $\mathcal{M}$ and the coordinates $x_i$ which we take to be $(u,v,x,r)$,
action of conformal gravity is defined by \begin{align}
    S_{CG}=\alpha_{CG}\int_{\mathcal{M}}d^4x \sqrt{-g} C^{\alpha}{}_{\beta\gamma\rho}C_{\alpha}{}^{\beta\gamma\rho} \label{cgac}
\end{align}
for $C^{\alpha}{}_{\beta\gamma\rho}$ Weyl squared term, $\alpha_{CG}$ dimensionless constant, and $g_{\mu\nu}$ conformally invariant metric. Equation of motion of the action (\ref{cgac}) is called Bach equation \begin{align}
    (\nabla^{\delta}\nabla_{\gamma}+\frac{1}{2}R^{\delta}{}_{\gamma})C^{\gamma}{}_{\alpha\delta\beta}=0.\label{bach} 
\end{align}
The equation is fourth order in derivatives, and as a subset, it contains solutions of Einstein gravity. We want to consider the pp-wave metric which solves (\ref{bach}). Similar global solution of (\ref{bach}) from \cite{Irakleidou:2016xot} showed that there can be interesting holography directly related to unitarity of the theory. 

\textbf{Ans\"atze 1 and 2.}
We consider metric of a form
\begin{align}
ds^2=\frac{f(r)}{h(r)}du^2+\frac{2}{h(r)}dudv+k(r)dr^2+k(r)dx^2\label{eq11}
\end{align}
which solves the Bach equation for
\begin{align}
k=\frac{c_2 e^{-c1 r} }{h(r)}&& \text{ and } && f= \frac{c_1 c_2-2c_3+rc_3 c_1}{c_1^3}+e^{-c_1 r}( c_4+rc_5).
\end{align}
Where one can recognise the $\frac{1}{h(r)}$ as a conformal factor. Due to conformal invariance of Bach equation each metric with arbitrary conformal factor is also a solution. To investigate symmetries of this solution we examine its Killing vectors (KV). For arbitrary $h(r)$
 conformal Killing equation $\nabla_{\mu}\xi_{\nu}+\nabla_{\nu}\xi_{\mu}=\frac{1}{2}g_{\mu\nu}\nabla_{\alpha}\xi^{\alpha}$ is satisfied by KVs of translations 
\begin{align}
\xi_x&=(0,0,1,0)\label{kvn1}, &&
\xi_u=(1,0,0,0) &&
\xi_v=(0,1,0,0),
\end{align}
while in the special case of conformally flat metric, when $c_1=1$, $c_3=0$, $c_5=0$  (ansatz 2), there are two additional KVs,
\begin{align}
\xi_1&=a_1\left(0,e^{\frac{1}{2}(u+r)}\cos\left(\frac{1}{2}x\right), -e^{\frac{1}{2}(u-r)}\sin\left(\frac{1}{2}x\right),-e^{\frac{1}{2}(u-r)}\cos\left( \frac{1}{2}x \right)\right) \label{kv1} \\
\xi_2&=a_1\left(0,-e^{\frac{1}{2}(u+r)}\sin\left(\frac{1}{2}x\right), -e^{\frac{1}{2}(u-r)}\cos\left(\frac{1}{2}x\right),e^{\frac{1}{2}(u-r)}\sin\left( \frac{1}{2}x \right)\right) \label{kv2}.
\end{align}
This indicates that the solution in that special case becomes a plane wave.
The KVS define commutation relations 
\begin{align}
    [\xi_x,\xi_2]&=\frac{1}{2}\xi_1, && [\xi_x,\xi_1]=\frac{1}{2}\xi_2,&& [\xi_u,\xi_2]=\frac{1}{2}\xi_2,&&
    [\xi_u,\xi_1]=\frac{1}{2}\xi_1 \label{comgl}
\end{align}
which can be recognized as two separate algebras. Redefining $\tilde{\xi}_x=2\xi_x$ the first two commutation relations in (\ref{comgl}) close Bianchi V algebra \cite{Popovych:2003xb}, while using $\tilde{\xi}_u=2\xi_u$ the latter two close the Bianchi VII algebra. 
Some examples of Bianchi Universes of the type V can be found in \cite{Khan:2017ziu}, and types IV, VI${}_h$, VII$_h$ in \cite{araujo}. For comparison to other Bianchi types, one can look at  type I in \cite{Demaret:1998dm}, type III in \cite{Reddy:2012na}, and type II, VIII, IX  in \cite{Saha:2011hd}.

The solution (\ref{eq11}) is Type N solution in the Petrov classification 
which we calculate using Mathematica program RGTC. 
Conformal gravity solutions, particularly of Petrov N type, have been studied in \cite{fiedler}.
The studies of the gravitational waves in quadratic curvature  gravity using Newman-Penrose formulation have been studied for Petrov D solutions in \cite{Baykal:2014exa}.

\section{Asymptotic Analysis}
To analyse holography of (\ref{eq11}) we transform coordinates $u\rightarrow aq-cy$ and $v\rightarrow bq+dy$, and take $h(r)=r^2$,  obtaining the metric
\begin{align}
  ds^2= \frac{1}{r^2}&(c_2 dr^2+ dq dy \left(2H(r)+2a^2 e^{-c_1r} \left( c_5 r-1\right)\right)\nonumber+dq^2 \left(H(r)+e^{-c_1r} \left(a^2 \left(c_5 r-1\right)-1\right)\right)\nonumber\\&+dy^2 \left(H(r)+e^{-c_1r} \left(a^2 \left(c_5 r-1\right)+1\right)\right)+c_2dx^2).\label{eq41}
\end{align}
 where $H(r)=a^2 \left(\frac{ c_3 r}{c_1^2}+\frac{ c_2}{c_1^2}-\frac{2 c_3}{c_1^3}\right)$. 
 The metric has a Rici scalar equal to $R=-\frac{3 \left(c_1^2 r^2+4 c_1 r+8\right)}{2 c_2}$ which can not take Ricci flat form by suitable choice of parameters. Ricci scalar is inversely proportional to $c_2$ which if sent to infinity,  would cause metric to diverge.
We expand (\ref{eq41}) to Fefferman-Graham (FG) form  $ds^2=\frac{1}{r^2}(dr^2+\gamma_{ij}dx^idx^j)$, for the $\gamma_{ij}$ metric at the boundary, and $r$ holographic coordinate. The metric at the boundary is expanded in the terms of the small perturbations  around $r=0$, such that  $\gamma_{ij}=\gamma_{ij}^{(0)}+\gamma_{ij}^{(1)} r +\gamma_{ij}^{(2)}r^2 +\gamma_{ij}^{(3)}r^3$, for $\gamma_{ij}^{(I)}$, $I=0,1,2,3$ matrices given in the expansion. 
When $b\to -\frac{a^2 (c_4+1)+1}{2 a}$, $d\to -\frac{1}{2} a (c_4+1)+\frac{1}{2 a}$ and $c\to -a$
we can choose the matrix $\gamma_{ij}^{(0)}=diag(-1,1,1)$ to be Minkowski metric, where the time coordinate is $q$. That defines the  $\gamma_{ij}^{(1)}$ matrix to be
\begin{align}
   \gamma^{(1)}_{ij}&= \left(
\begin{array}{ccc}
 \left(\frac{c_1-c_3}{c_1^2}+c_5\right) a^2+c_1 & \frac{a^2 \left(c_5 c_1^2+c_1-c_3\right)}{c_1^2} & 0 \\
 \frac{a^2 \left(c_5 c_1^2+c_1-c_3\right)}{c_1^2} & a^2 \left(\frac{c_1-c_3}{c_1^2}+c_5\right)-c_1 & 0 \\
 0 & 0 & 0 \\
\end{array}
\right).\label{g1}
\end{align}
This matrix can be compared with $\gamma_{ij}^{(1)}$ term from the FG expansion for Manheimm-Kazanas-Riegert (MKR) solution \cite{Riegert:1984zz,Grumiller:2010bz,Grumiller:2013mxa}. The MKR solution is different from (\ref{eq41}), however we 
can use its properties to better understand the meaning of parameters which appear in our case. 
In FG expansion of MKR solution, $\gamma_{ij}^{(1)}$ matrix depends entirely on the term that describes Rindler acceleration. If we are drawing analogous conclusion in (\ref{g1}), this role is played by the combination of $c_1,c_3,c_5$ parameters. Parameter $a$ from (\ref{eq41}) can be absorbed in the coordinate, so it does not carry physical meaning. $\gamma_{ij}^{(2)}$ matrix of MKR does not show explicit dependence on mass parameter when Rindler acceleration parameter vanishes, so we consider matrix $\gamma_{ij}^{(3)}$ 
\begin{align}
\gamma_{ij}^{(3)} = \left(
\begin{array}{ccc}
 c_1^3+a^2 \left(3 c_5 c_1^2+c_1-2 c_3\right) & a^2 \left(3 c_5 c_1^2+c_1-2 c_3\right) & 0 \\
 a^2 \left(3 c_5 c_1^2+c_1-2 c_3\right) & a^2 \left(3 c_5 c_1^2+c_1-2 c_3\right)-c_1^3 & 0 \\
 0 & 0 & 0 \\
\end{array}
\right).\label{gam3}
\end{align}
If Rindler parameter in MKR solution is zero, it's $\gamma_{ij}^{(3)}$ matrix  is given solely in terms of the mass parameter \cite{Grumiller:2013mxa}. This implies that combination of parameters in (\ref{gam3}) carries physical meanings of mass and Rindler acceleration.  
Now, one can compute the holographic stress energy tensors of (\ref{eq41}) $\tau_{ij}$ and $P_{ij}$, by inserting $\gamma_{ij}^{(1)},\gamma_{ij}^{(2)}$ and $\gamma_{ij}^{(3)}$ in the $\tau_{ij}$   and $P_{ij}$ \cite{Grumiller:2013mxa}. The stress energy tensor $\tau_{ij}$ and PMR are  given by 
\begin{align}
\tau_{ij}&=  a^2 c_1^2 c_5  \left(
\begin{array}{ccc}
 1 & 1 & 0 \\
 1 &1 & 0 \\
 0 & 0 & 0 \\
\end{array}
\right),
&&    P_{ij}= \frac{a^2 \left(c_1^2 c_5-c_3\right)}{c_1}\left(
\begin{array}{ccc}
1 & 1 & 0 \\
 1 & 1 & 0 \\
 0 & 0 & 0 \\
\end{array}
\right)\label{pij}
\end{align} respectively. The Ward identity of CG is satisfied with $2\tau_{ij} \gamma^{(0)ij}+P_{ij}\gamma^{(1)ij}=0$. 
For $\chi^{(0)k}$ asymptotic KV, the current
$J^i=Q^{ij}\chi^{(0)}_j $
is conserved $\mathcal{D}_iJ^i=0$. The corresponding charge $Q_{ij}=2\tau_{ij}+P_{ik}\gamma^{(1)k}{}_{j}+P_{ki}\gamma_j^{(1)k}$
for  (\ref{eq41}) 
\begin{align}
 Q_{ij}= 2 a^2 c_3\left(
\begin{array}{ccc}
 1 & 1 & 0 \\
 1 & 1 & 0 \\
 0 & 0 & 0 \\
\end{array}
\right)\label{qij}
\end{align}
 is given in terms of $c_3$.  For the specific case of the metric 
 \begin{small}
 \begin{align}
   ds^2=\frac{1}{r^2}\left( c_2 dr^2+ 2a^2\left( c_2- e^{-r}\right)dq dy +dq^2 \left(a^2 c_2-\left(a^2+1\right) e^{-r}\right)+dy^2 \left(c_2(a^2+1) +\left(1-a^2\right) e^{-r}\right)+c_2dx^2\right),\label{specrot}
 \end{align}
 \end{small}
 with five KV's (\ref{kvn1})-(\ref{kv2}), when $c_1=1,c_3=c_5=0$, 
  the stress tensors $\tau_{ij}= 
P_{ij}=Q_{ij}=0$ and charges exactly vanish,  as we can see from the general expressions for the stress energy tensors (\ref{pij}), and charge (\ref{qij}). This is expected from the highly symmetric solution, which also describes conformally flat metric.
That choice of parameters still does not give a metric which satisfies Einstein equations.  Interestingly, there is no such choice of parameters which would make solution (\ref{eq41}) satisfy Einstein vacuum equation.

It is well known that conformal gravity is non-unitary. In \cite{Ghodsi:2014hua} the non-unitarity of conformal gravity was shown via two-point correlation functions. It manifests through the negative sign of the correlation function with PMR. Since PMR is zero for the (\ref{specrot}) that issue is avoided.
However, we have also vanishing $\tau_{ij}$ and a vanishing $Q_{ij}$. The coformal flatness means that the entropy of the solution (visible also from Weyl squared) is going to be zero. From (\ref{pij}) we see that $P_{ij}$ will vanish for $c_3=c_5$ when $c_1=1$. This choice of metric has only three global KVs but it is not conformally flat. The stress energy tensor $\tau_{ij}$ and charge $Q_{ij}$ are visible from (\ref{pij}) and (\ref{qij}) and they do not vanish. 
This appears as well in \cite{Grumiller:2013mxa} for the example of rotating black hole \cite{Liu:2012xn}, where $\gamma^{(1)}_{ij}$ does not vanish while $P_{ij}$ vanishes.
If in our example (\ref{eq41}), we demand $\gamma^{(1)}_{ij}$ to be zero, that implies $P_{ij}$ (\ref{pij}) is as well zero.
The vanishing of $\gamma_{ij}$ however, does not always automatically imply vanishing of the $P_{ij}$. On the example of the pp-wave solution \cite{Irakleidou:2016xot},
\begin{align}
    ds^2=\frac{1}{r^2}(dr^2+(-1+f(r))dx^2+2f(r)dxdy+(1+f(r))dy^2+dz^2)
\end{align}
where $f(r)=c_1+c_2 r+c_3r^2+c_4r^3$
 one can fix the $\gamma^{(1)}_{ij}$ to be zero (setting $c_2=0$), without affecting the $P_{ij}$ which becomes defined solely by the $\gamma^{(2)}_{ij}$ and $c_3$.

The charges that we calculated, express stress energy tensors and charge in a sense of \cite{Hollands:2005ya} which differ from the other such charges (for example those defined by Hamiltonian method as in \cite{Caprini:2018oqe}) only by a "constant offset" determined by boundary fields alone. The algebra generated by the charges in conformal gravity is equivalent to the Lie algebra of the transformations preserving boundary conditions, i. e. asymptotic symmetry algebra \cite{Irakleidou:2014vla}.

{\bf Asymptotic symmetry algebra} (ASA) for conformal gravity has been studied in \cite{Irakleidou:2016xot}
.To obtain ASA for (\ref{eq41}) one has to study the expansion of the the  conformal Killing equation (CKE) 
in the coordinate $r$. 
 The metric at the boundary was chosen to be   $\gamma_{ij}^{(0)}=diag(-1,1,1)$ Minkowski metric, which leads to full conformal algebra in the leading order of expansion of CKE. The subleading order of CKE equation \cite{Irakleidou:2016xot} 
 \begin{align}
     \pounds_{\xi^{(0)}}\gamma_{ij}^{(1)}=\frac{1}{3}\mathcal{D}_k\xi_{(0)}^k\gamma_{ij}^{(1)}
 \end{align} defines the  
  ASA 
  \begin{align}
     [\xi_t,\xi_{L_1}]&=\xi_x,&& \left[\xi_t,\xi_{L_2}\right]=\xi_y+\frac{1}{2}\xi_t, && [\xi_y,\xi_{L_1}]=-\xi_x, && [\xi_y,\xi_{L_2}]=\frac{1}{2}\xi_{t}+\xi_y, \\ [\xi_x,\xi_{L_1}]&=\xi_t+\xi_y, && [\xi_x,\xi_{L_2}]=\xi_x,&&
     [\xi_{L_1},\xi_{L_2}]=\frac{1}{2}\xi_{L_1}&&
  \end{align}
  with KVs 
  \begin{align}
      \xi_t&=(1,0,0), && \xi_x=(0,0,1) && \xi_y=(0,1,0) \\
      \xi_{L_1}&=(x,x,t-y) && \xi_{L_2}=(t+\frac{1}{2}y,\frac{1}{2}t+y,x)&&.
  \end{align}
 The ASA is unaffected by the choice of the parameters $c_i$, $i=1,..,5$, and it is equal for each of the special cases of the solution (\ref{eq41}). It belongs to the ASA $a_{5,4}^a$ for $a=\frac{1}{2}$ from \cite{Patera:1976my}. The classes of five dimensional ASAs have been encountered in CG \cite{Irakleidou:2016xot}.

\textbf{Applications of the metric.}
If we look at the metric as Einstein solution with an additional matter, we can have following considerations.
After the transformation of coordinates $e^{c_1 r/2}\rightarrow z$ and choice for conformal factor $h(r)=1/(4z^2\ln(z))$ and $c_1=1$, 
one can relate the metric (\ref{eq11}) with the metric  
\begin{align}
ds^2=2dudv+f(z)du^2+z^2dx^2+dz^2\label{metl}
\end{align}
which solves Bach equation for 
$
f(z)=(\frac{1}{4}-2c_3)z^2+c_4+2c_5\log{z}+2z^2c_3\log{z}, 
$ 
 and $c_2=\frac{1}{4},x=2\tilde{x}.$ Where we omit "$\sim $" for simplicity. 
This solution is similar to the metric considered in \cite{Papadopoulos:2002bg}. 
There, the metric \begin{align}
 ds^2=2dudv-\lambda(u)x^2du^2+dx^idx^i\label{at}   
\end{align} was studied for the propagation of string modes and  first-quantized point-particle  
in this time-dependent background. Where $dx^idx^i=dx^2+dz^2$  Euclidean metric. 

\textbf{Ansatz 3.}  Generalization of the metric (\ref{metl}) by multiplying  $f(z)$ with $\lambda(u)$ does not influence the solvability of Bach equation. One may wonder if further simple generalizations are possible.
We consider the metric 
\begin{align}
    ds^2=2dudv+f(u,x,z)du^2+dx^2+dz^2\label{eq56}
\end{align}
where we immediately crossed from cylindrical to Euclidean coordinates.
The generalization by introducing the dependency $z$ so that $f(z)\rightarrow f(x,z)$ in (\ref{metl}) leads to the fourth order equation which can be decomposed into  $(-\partial_z+i\partial_x)^2(\partial_z+i \partial_x)^2f(x,z)=0$. The solution is \begin{align}
f(x,z)\rightarrow (d_1+d_2x+d_3z) f_1(-i x+z) + (d_4+d_5x+d_5z)f_2(ix+z) \label{solim}
\end{align} 
They can obviously become of the interesting form for trigonometric and exponential functions (we will mention specific cases later). 
\footnote{We can bring the solution (\ref{uxz}) to the form of the metric studied in \cite{araujo}, by considering the transformation 
$x\rightarrow i x_1+\frac{i}{2}x_2$  and $z\rightarrow x_1+\frac{1}{2}x_2$. 
The obtained metric reads  
\begin{align}
    ds^2=H_1(x_1,x_2)du^2+2dudv-2dx_1dx_2\label{eq61}
\end{align}
The general form of the solution for 
obtained from the Bach equation would lead to $H_1=f_1(x_2)+x_1f_2(x_1)+f_3(x_1)+x_2f_4(x_1)$. Only keeping $f_1$ and $f_3$ satisfies Einstein solution and can be cast into form studied in \cite{araujo}, while CG solution involves all four functions.
}

The assumption $f(x,z)\rightarrow f(u,x,z)$ will  let us write the solution as
\begin{align}
 f(u,x,z)&=(d_1+d_2x+d_3z)f_1(u,-i x+z) + (d_4+d_5x+d_5z)f_2(u,ix+z) 
 \label{uxz}.
\end{align}
Where we took into account that each of the functions depending on $(x,z)$ can be multiplied by arbitrary function of u.
The metric (\ref{eq56}) is completely equal to the ansatz metric in \cite{Peres:1959mm} after appropriate transformation of the coordinates. The statement that Einstein equations in vacuum are satisfied for every harmonic function $f$ which is a function of $x$ and $z$, whatever was the dependence on $u$, is now generalized. The Bach equations in vacuum are satisfied for every harmonic function $f$ which is a function of $x$ and $z$ multiplied by arbitrary function of $u$ and by the ($d_1+d_2 x+d_3 z$) or ($d_4+d_5 x+d_6 z$) for $d_1,d_2,d_3,d_4,d_5,d_6$ which are arbitrary, or defined, depending on the function we want to express.
For example 
\begin{itemize}
    \item $f(u,x,z)=2(d_1+2_2 x+d_3z)\arctan\left(\frac{z}{x}\right)$, for $f_1=i \log(x-iz)$, $f_2=i\log(x-iz)$, and $d_1=-d_4,d_2=-d_5$ and $d_3=-d_6$. This is a term from rotation of metric analogous to (\ref{metl})  from cylindrical to Euclidean 
    \item $f(u,x,z)=b_2(x^2+z^2)\log(x^2+z^2)$, for $f_1=(x-iz)\log(x-iz)$, $f_2=(x+iz)\log(x+iz)$, and $d_3=id_2,d_5=d_2,$ and $d_6=-id_2$. This is an additional term in generalization of plane wave metric (\ref{at}), \cite{Papadopoulos:2002bg} .
\end{itemize}

The above metric (\ref{eq56}) conserves only one KV, that is $\partial_v$.
If we want to consider the solution (\ref{eq56}) (with $f(u,x,z)$ from (\ref{uxz})) as the background for the string propagation, we are going to need to transform it to Rosen coordinates, following the procedure of  \cite{Papadopoulos:2002bg}.
For functions $f_1$ and $f_2$ such that the metric (\ref{eq56}) is $l(u)(x^2+z^2)$ 
reduces to case studied earlier in \cite{Papadopoulos:2002bg}. Here we focus on the study of the situation when $f_{1},f_2$
are more general. 
In the asymptotic analysis we are going to keep writing $f(u,x,z)$ in terms of functions dependent on $x+iz$ and $x-iz$, however this is only to keep the functional dependence, while one needs to keep in mind that for each specific case metric needs to be of course real.

{\bf Asymptotic analysis.}
Transformations
$u\rightarrow -\frac{1}{2a_3}(q+y)$ , $v\rightarrow a_3 (q - y)$ where we can choose for simplicity $a_3=-\frac{1}{2}$, lead from (\ref{eq56}) and (\ref{uxz}) to
\begin{align}
ds^2=&\frac{1}{z^2}( (-1+g_1(q+y,x,z)) dq^2+2(g_1(q+y,x,z))dydq  \nonumber \\ &+ (1+g_1(q+y,x,z)) dy^2+dx^2+dz^2 ) \label{eq319}
\end{align}
for
$ g_1(q+y,x,z)= (d_1+d_2x+d_3z) f_1(q+y,x-i z)+(d_3+d_5x+d_6z) f_2(q+y,x+i z).$  The Ricci scalar of the metric is -12, while it is zero for conformally invariant metric  (\ref{eq56}) multiplied by $z^2$, and the KV is $\partial_v$. FG expansion of the metric (\ref{eq56}) in $z$ coordinate, done analogously to the expansion of (\ref{eq41}), requires $d_1=d_3, d_5=d_2$ and $f_1(q+y,x)=-f_2(q+y,x)$, and it results with the $\gamma^{(1)}_{ij}$ matrix
\begin{align}
  \gamma^{(1)}_{ij}=\left(
\begin{array}{ccc}
 h_1 & h_1 & 0 \\
 h_1 & h_1 & 0 \\
 0 & 0 & 0 \\
\end{array}
\right), && \gamma_{ij}^{(2)}=\left(
\begin{array}{ccc}
 0 & 0 & 0 \\
 0 & h_2 & 0 \\
 0 & 0 & 0 \\
\end{array}
\right)
\end{align}
for
$
    h_1=h_1(q+y,x)=\left.(d_6-d_3\right)f_2(q+y,x)-2 i \left(d_5 x+d_4\right) f_2{}^{(0,1)}(q+y,x)
    $ and $
h_2=h_2(q+y,x)=\frac{1}{3} i \left(d_3+d_6\right) f_2{}^{(0,3)}(q+y,x).
$
From $h_1$ and $h_2$ and comparison to the (\ref{g1}) and (\ref{gam2}) respectively, we can see that Rindler parameter and mass are given by combination of the parameters $d_6,d_3,d_5$ and $d_4$.
Expressing the 
stress tensors in terms of the function $f_2(q+y,x)$ and its derivatives, allows to see functional dependence from the metric directly in the response functions and charge.
Stress tensor $\tau_{ij}$ is given by
\begin{align}
 \tau_{ij}=  \left(
\begin{array}{ccc}
 h_1h_2+h_3 & -\frac{6}{7}h_1h_2 & 0 \\
 -\frac{6}{7}h_1h_2 & -\frac{3}{4}h_1h_2+\frac{3}{2}h_3 & 0 \\
 0 & 0 &\frac{6}{5} h_1h_2+\frac{6}{5}h_3 \\
\end{array}
\right)\label{tijsc}
\end{align}
for $h_3=h_3(q+y,x)=\frac{1}{60} \left(5 \left(d_6-d_3\right) f_2{}^{(0,4)}(q+y,x)-2 i (d_4+d_5 x) f_2{}^{(0,5)}(q+y,x)\right)$ and 
\begin{align}
    P_{ij}=-\frac{1}{3} h_2,\label{pijsc}
\end{align}
while definition of charge is given in terms of $h_1,h_2$ and $h_3$, see appendix (\ref{charge2}).  
It is important to notice that for this metric, one can choose $d_6=d_3$ and $d_5=d_4=0$ which will lead to vanishing of the $\gamma_{ij}^{(1)}$, while the PMR tensor will not vanish. This is due to 
proporcionality of $P_{ij}$ to $\gamma_{ij}^{(2)}$ and non-vanishing $\gamma_{ij}^{(2)}$. This is specific property of the solution, observed only for the pp-wave solution solution in \cite{Irakleidou:2016xot}.

By choosing $g_1(q+y,x,z)$ we can set the metric to be
\begin{align}
    ds^2=\frac{1}{z^2}\left( -(1+4zf(q+y,x,z))dq^2- 8zf(q+y,x,z) dqdy+(1-4zf(q+y,x,z))dy^2+dx^2+dz^2\right)\label{cossol}
\end{align}
for $f(q+y,x,z)=b\left(5 x^4-10 z^2 x^2+z^4\right) \cos \left(q+y\right).$ The stress tensor 
$$  \tau_{ij}= diag\left(
 -8b \cos (q+y),
 -16b\cos (q+y),
 8 b\cos (q+y) 
\right)$$
is defined from only non-vanishing matrices in the FG expansion, $\gamma_{ij}^{(1)}$ and $\gamma_{ij}^{(3)}$ given in the appendix (\ref{gam1spec}). Since $\gamma_{ij}^{(2)}$ is zero, the partially massless response tensor, 
$P_{ij}$ vanishes, which results with $Q_{ij}=2\tau_{ij}$.  For the only parameter $b$ which we have here, we can conclude to have a role similar as a Rindler parameter.

{\bf The asymptotic symmetry algebra} for this metric is three dimensional, consisting of KVs $\xi_1=(0,0,2a_1),\xi_2=(-a_2,a_2,0), \xi_3=(-\frac{x}{2},\frac{x}{2},-\frac{a_3}{2} (q+ y))$ which close the algebra $[\xi_1,\xi_2]=\frac{a_1}{a_2}\xi_3$, which is
for $a_1=a_2=1$, Bianchi II algebra, also called Heisenberg-Weyl algebra.

Metric (\ref{eq319}) can be reduced to a Ricci flat metric via conformal transformation, multiplying it with $z^2$. That metric can be transformed to a flat metric which one can naturally write in the Rosen form.

\section{Background for wave propagation}

\textbf{Comment on metric as backgroud for studying string propagation.}
To see if the metrics (\ref{eq56}) for $f$ given in (\ref{uxz}), could be considered as a background for studying the string propagation we follow the procedure in \cite{Papadopoulos:2002bg}.
To embed the metric into string theory one needs to compensate the non-zero Ricci tensor $R_{uu}$ of the metric with the contribution from other background fields. 
Without restricting parameters, the $R_{uu}$ of our metric (\ref{eq56}) gives
\begin{align}
R_{uu}=&-\left(d_1+i d_2\right) f_1{}^{(0,1)}(q,x+i z)-\left(d_4-i d_5\right) f_2{}^{(0,1)}(q,x-i z)\end{align}
which indicates that choosing $d_1=-id_2$, $d_4=id_5+\frac{1}{2}$, and $f_2=(x-iz)\lambda(u)$ will lead to the similar $R_{uu}$ as in \cite{Papadopoulos:2002bg}, i.e. $R_{uu}=-\frac{1}{2}\lambda(u)\label{rict}.$ 
This implies that the metric-dilaton cosmological background as a pp-wave analog, can be given by 
\begin{align}
ds^2=&2dudv+F(u,x,z)du^2+dx^idx^i, & &\phi=\phi(u)\label{met3}
\end{align}
for \begin{align}
    F(u,x,z)=\left(d_0+d_2 (z-i x)\right) f_1(q,x+i z)+\lambda (u) (x-i z) \left(d_5 (z+i x)+d_3+\frac{x}{2}\right)  \label{func}
\end{align}
where $\phi$ is $u$ dependent dilaton field.
The conformal invariance condition for Ricci tensor 
(\ref{rict}) is \cite{Papadopoulos:2002bg}  $R_{uu}=-2\partial_u^2\phi$ and it leads to $\phi''(u)=\frac{d}{2}\lambda(u)$. For the special case $\lambda=\frac{k}{u^2}$ the family of pp-wave backgrounds that one obtains is
\begin{align}
    ds^2=&2dudv-\frac{1}{2} \frac{k}{u^2} F(u,x,z)du^2+dx^idx^i\label{eq513}\\
    e^{2\phi}=&u^{kd}e^{2u} \label{eq514}
\end{align}
Before one can approach to the string theory quantization in the metric-dilaton background, one usually considers quantum theory of a propagation  of scalar relativistic particle in that background.
Covariant quantization of a relativistic particle gives Klein-Gordon equation for the considered background.  The covariant-quantization of this case can be related to the quantization on the light-cone gauge due to $V=\partial_v$ isometry of the background. In the case of the background (\ref{at}) and $\lambda(u)=\frac{k}{u^2}$ \cite{Papadopoulos:2002bg} this light-cone quantization reduces to a problem of harmonic-oscillator with time dependent frequency.

For the $\Phi$ a space time-field which represents a massive scalar-string mode and $\tilde{\Phi}=e^{-\phi}\Phi$, one can write the string free field Klein-Gordon equation as \cite{Papadopoulos:2002bg}
\begin{align}
    \left(-\frac{1}{\sqrt{G}}\partial_{\mu}(\sqrt{G}G^{\mu\nu}\partial_{\nu})+m^2 \right)\tilde{\Phi}=0.
\end{align}
The equation depends on the string-frame metric $G^{\mu\nu}$, and not on the dilaton $\phi$.

The general solution to the KG equation which is real is \cite{Papadopoulos:2002bg} $\tilde{\Phi}=\sum_k[\alpha_k\varphi_k(x)+\alpha_k^*\varphi_k(x)]$ for $\{\varphi_k,\varphi^*_k\}$ a set of special solutions. The basis $\varphi_x,\varphi^*_k$ has an explicit form which depends on the choice of coordinates and boundary conditions, and it has normalisation given in the appendix B. 

The choice of coordinates that one can consider are for example conformally-flat coordinates and Brinkmann coordinates.  In conformally-flat coordinates,  the function  (\ref{func}) of the metric (\ref{met3}) becomes $F(u,x,z)=\frac{1}{4} \lambda (q) (x-i z) \left(4 d_3+x+i z\right)+d_2 (z-i x) \left(\lambda _2(q)+\lambda _3(q) (x-i z)\right)+d_0$ based on the condition of conformal-flatness and requirement that Ricci tensor is (\ref{rict}). 

In the Brinkmann coordinates, the massless KG equation  for the background (\ref{eq513}), (\ref{eq514}) reads
\begin{align}
    (2\partial_u\partial_v+F(u,x,z)\partial_v^2+\delta^{ij}\partial_i\partial_j)\tilde{\Phi}_0=0\label{eq517}
\end{align}
where $\tilde{\Phi}_0=\int dp_v e^{ip_vv}\psi(u,x,z;p_v)$. In the case when $F(u,x,z)$ in (\ref{eq513}) reduces to $x^ix_i$, the problem considers harmonic  oscillator with time-dependent frequency. Otherwise one needs to consider modified harmonic-oscillator problem, and transformation of metric (\ref{eq513}) to Rosen coordinates. 

To  study the metrics within a framework of conformal gravity, one would have to require that $R_{uu}$ vanishes, since there is no notion of dilaton-metric backround for the conformal gravity. 
From the equation (\ref{eq517}) one could see the form of the KG equation obtained for each of these  metrics in Brinkmann form. Limiting the solution to Einstein solution and asking that $R_{uu}=0$ might lead to tractable problems. In this case the pp-wave metrics that would be considered would be exclusively those defined by metrics in \cite{Peres:1959mm}.







\section{Conclusion}

We have studied the pp-wave solution of conformal gravity and its symmetries. The most general form of the solution admits three translational Killing vectors, while choosing specific parameters the symmetries can be increased to five KVs. 
Via asymptotic analysis we calculate holographic stress energy tensors of conformal gravity. The most symmetric solution has  both stress energy tensors vanishing, as well as vanishing Weyl and  charge. For the specific choice of parameters we find vanishing PMR for a metric which is not conformally flat, and does not have vanishing charge or Brown-York stress tensor. In other words, zero PMR does not imply that the global solution becomes Einstein solution.  The interesting thing is that  non-unitarity of the conformal gravity manifests through PMR, when PMR is zero, this is not the case, which renders this solution important.
The second pp-wave solution we study is also most general solution of its respective form, and it is a generalisation of the pp-waves in Einstein gravity studied in \cite{Peres:1959mm}. The holography of this solution shows that one can have vanishing subleading term in the FG expansion and non-vanishing PMR, which makes it a first example of its kind.  

We also considered possible application of the studied pp-wave metrics. In the future studies it would be interesting to use these metrics as a background for the certain calculations as string propagation. It would be also interesting to use concrete example of the metric on the calculation as \cite{Bender:2007wu} and further see if it could be written directly using partially-massless response function.


\section{Acknowledgments}
The author would like to thank Arkady Tseytlin and Daniel Grumiller for the discussions and comments, and Toby Wiseman, and Matthew Roberts for discussions. I.L. was supported by the FWF Schr\"odinger grant J 4129-N27.

\section{Appendix A}

Matrix $\gamma_{ij}^{(2)}$ in the FG expansion of (\ref{eq41}) is

\begin{align} \gamma_{ij}^{(2)}&=\left(
\begin{array}{ccc}
 a^2 \left(\frac{2 c_3}{c_1}-2 c_1 c_5-1\right)-c_1^2 & -\frac{a^2 \left(2 c_5 c_1^2+c_1-2 c_3\right)}{c_1} & 0 \\
 -\frac{a^2 \left(2 c_5 c_1^2+c_1-2 c_3\right)}{c_1} & \left(\frac{2 c_3}{c_1}-2 c_1 c_5-1\right) a^2+c_1^2 & 0 \\
 0 & 0 & 0 \\
\end{array}
\right).\label{gam2}
\end{align}
The charge defined by the (\ref{tijsc}) and (\ref{pijsc}) is
\begin{align}
    Q_{ij}=\left(
\begin{array}{ccc}
-\frac{3}{4}h_2h_1+\frac{3}{2}h_3 & -\frac{3}{11}h_2h_1 & 0 \\
-\frac{3}{11}h_2h_1 & -\frac{1}{4}h_2h_1 +h_3 & 0 \\
 0 & 0 & \frac{3}{5}h_2h_1 +h_3 \\
\end{array}
\right).\label{charge2}
\end{align}
The $\gamma_{ij}^{(k)}$ for k=1,2,3 matrices for  the metric (\ref{cossol}), are given by
\begin{align}
    \gamma_{ij}^{(1)}&=-20 bx^4 \cos (q+y)\left(
\begin{array}{ccc}
 1 & 1 & 0 \\
 1  & 1 & 0 \\
 0 & 0 & 0 \\
\end{array}
\right), && \gamma_{ij}^{(3)}=-24 b\cos (q+y)\left(
\begin{array}{ccc}
 0 & 0 & 0 \\
 0  & 1 & 0 \\
 0 & 0 & 0 \\
\end{array}
\right).  \label{gam1spec}
\end{align}

\section{Appendix B} 

Normalisation of the specific solutions is \begin{align}
    \int d^{D-1}x\sqrt{-G}G^{0\mu}(\varphi_k^*\partial_{\mu}\varphi_k'-\varphi_k'\partial_{\mu}\varphi_k^*)=-i\delta_{kk'} \label{norm}
\end{align}

\section{Appendix C}
We consider the geodesics of the metric (3.1) after the transformation $e^{c_1 r/2}\rightarrow z$ and for the choice of parameters $c_5=\frac{1}{2}, c_3=0,a_2=1/2$ while the other parameters are set to 1. The geodesic equations lead to
\begin{align}
   & u''(s)-\frac{u'(s) z'(s)}{z(s) \log (z(s))}=0\\
    &\frac{u'(s) \log \left(z(s)\right) \left(2 z(s) y'(s)+\left(z(s)^2+2\right) z'(s)\right)-2 v'(s) z'(s)}{2 z(s) \log \left(z(s)\right)}+v''(s)=0\\
    &\frac{1}{8 z(s) \log (z(s))}(8 u'(s) v'(s)+u'(s)^2 \left(4 (y(s)+2)+z(s)^2-2 z(s)^2 \log (z(s))\right)\nonumber \\& +4 z(s)^2 y'(s)^2 (1-2 \log (z(s)))-4 z'(s)^2)+z''(s)=0\\
    &-\frac{u'(s)^2}{2 z(s)^2}+y''(s)+\frac{y'(s) z'(s) (2 \log (z(s))-1)}{z(s) \log (z(s))}=0
\end{align}
They can be solved numerically for the set of initial conditions
which are given under the images, 
\begin{figure}[!ht]
\begin{center}
\includegraphics[width=18cm]{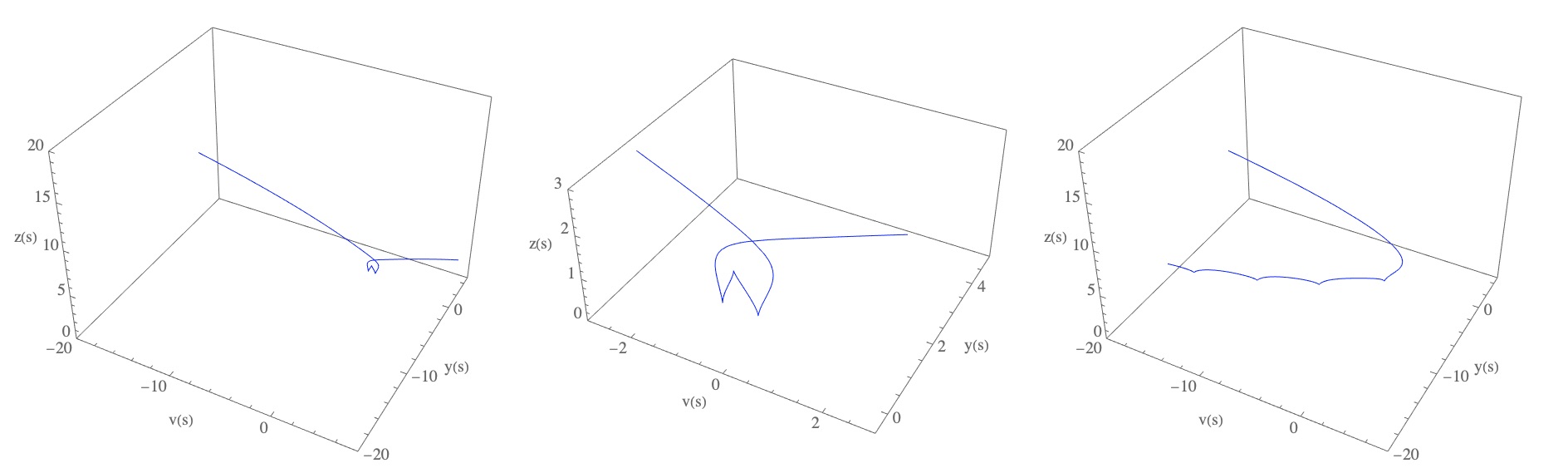}
\end{center}
\caption{Initial conditions of the two left graphics are $y(0)=0,z(0)=2,v(0)=0,z'(0)=0,y'(0)=0,v'(0)=0,u(0)=0,u'(0)=1$, while the initial conditions of the right graphics are $y(0)=0,z(0)=2,v(0)=0,z'(0)=1,y'(0)=2,v'(0)=0,u(0)=0,u'(0)=1$}\label{geo1}
\end{figure}

%
\bibliographystyle{unsrt}
\bibliography{bibc.bib}

\end{document}